# Moiré Fringes in Conductive Atomic Force Microscopy

L. Richarz,[1] J. He,[1] U. Ludacka,[1] E. Bourret,[2] Z. Yan,[2,3] A.T.J. van Helvoort,[4] and D. Meier[1]

[1]*Department of Materials Science and Engineering, NTNU Norwegian University of Science and Technology, Trondheim, Norway*
[2]*Materials Sciences Division, Lawrence Berkeley National Laboratory, Berkeley, CA, USA*
[3]*Department of Physics, ETH Zurich, Zurich, Switzerland*
[4]*Department of Physics, NTNU Norwegian University of Science and Technology, Trondheim, Norway*

(*Electronic mail: dennis.meier@ntnu.no)

(Dated: 3rd February 2023)

Moiré physics plays an important role for the characterization of functional materials and the engineering of physical properties in general, ranging from strain-driven transport phenomena to superconductivity. Here, we report the observation of moiré fringes in conductive atomic force microscopy (cAFM) scans gained on the model ferroelectric Er(Mn,Ti)O$_3$. By performing a systematic study of the impact of key experimental parameters on the emergent moiré fringes, such as scan angle and pixel density, we demonstrate that the observed fringes arise due to a superposition of the applied raster scanning and sample-intrinsic properties, classifying the measured modulation in conductance as a scanning moiré effect. Our findings are important for the investigation of local transport phenomena in moiré engineered materials by cAFM, providing a general guideline for distinguishing extrinsic from intrinsic moiré effects. Furthermore, the experiments provide a possible pathway for enhancing the sensitivity, pushing the resolution limit of local transport measurements by probing conductance variations at the spatial resolution limit via more long-ranged moiré patterns.

## I. INTRODUCTION

Moiré fringes form when two periodic patterns of lines or dots are superposed onto each other. This interference effect pervades all of science, including mathematics, physics, and medical diagnostics. In scanning-based microscopy, moiré effects are well-established phenomena and have been reported for different techniques, such as scanning electron microscopy (SEM)[1,2], scanning transmission electron microscopy (STEM)[3] and atomic force microscopy (AFM)[4]. Here, so-called scanning moiré patterns arise due to a combination of the applied raster scanning and periodic properties intrinsic to the sample. The emergent scanning moiré fringes carry important information about the material under investigation and have been used, for example, to detect strain fields[5–7], interfaces[8], and nanoscale deformations[4], representing a viable pathway for improving the resolution of the applied technique.[3]

Recently, going beyond the characterization of material properties, moiré effects are attracting increasing attention for the engineering of electronic responses. In contrast to scanning-related phenomena, moiré engineering relies on the interference of two or more intrinsic structural or electronic properties, giving rise to new physical effects. Fascinating examples range from superconductivity in graphene[9] to strain-driven modulations of the electronic conductivity and ferromagnetism of La$_{0.67}$Sr$_{0.33}$MnO$_3$ thin films on LaAlO$_3$ substrates[10]. As SEM-, STEM-, and AFM-based microscopy methods are frequently applied for investigating the emergent moiré physics, i.e., techniques that themselves can give rise to moiré effects, it is crucial to carefully distinguish between extrinsic (probe-related) and intrinsic (sample-related) phenomena.

Here, we study scanning moiré effects in conductive AFM (cAFM), that is, a scanning probe technique widely used to map local transport phenomena with nanoscale spatial resolution. Using the ferroelectric semiconductor Er(Mn$_{0.998}$,Ti$_{0.002}$)O$_3$ as an instructive example (referred to as Er(Mn,Ti)O$_3$ in the following), we demonstrate the formation of moiré fringes in cAFM conductance maps. We find that the moiré fringes change in response to variations in the scan angle and the density of measurement points. The data shows that they arise from an interplay of the AFM scanning parameters and the electronic properties of the material, identifying the measured conductance variations as a scanning moiré effect. The work expands previous topographic AFM-based studies towards transport measurements, giving guidelines for the study of moiré effects by cAFM and new possibilities for the nanoscale characterization of emergent electronic properties in functional oxides in general.

## II. SPATIAL VARIATIONS IN ELECTRONIC TRANSPORT

To record conductance maps, we apply conventional cAFM measurements using a commercial atomic force microscope (Asylumn Research, Cypher ES Environmental AFM). Fig. 1(a) shows a cAFM scan gained on an Er(Mn,Ti)O$_3$ single crystal with a voltage of 3 V applied to the back-electrode. The crystal is oriented by Laue diffraction and cut such that the polarization direction ($P \parallel c$) is perpendicular to the sample surface (out-of-plane polarization)[11]. In the cAFM scan, several bright lines are visible, indicating a locally enhanced conductance. These lines correspond to 180° domain walls separating domains of opposite polarization ($+P$ and $-P$)[12,13]. The enhanced conductance at these domain walls originates from an accumulation of oxygen interstitials, which has been studied in detail in refs.[14,15]

An additional modulation in conductance is visible in the cAFM data, forming an extended wave-like pattern throughout the imaged region. This pattern has a periodicity of about 100-200 nm and an amplitude of 5-10 pA as shown in Fig. 1(b). Within the different domains, both loop-shaped and curved lines are observed, which consistently change their curvature at the position of the domain walls.

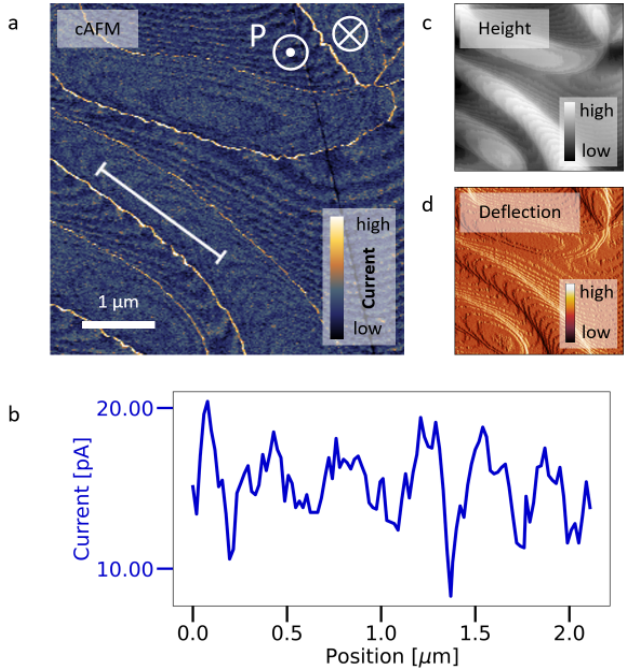

Figure 1. **Moiré pattern on Er(Mn,Ti)O$_3$.** (a) cAFM image recorded on an Er(Mn,Ti)O$_3$ sample with out-of-plane polarization. In addition to the enhanced current at the ferroelectric domain walls relative to the domains, a more subtle wave-like current variation is observed. (b) Current profile along the white line marked in (a), revealing a periodicity of approximately 250 nm. (c) Height and (d) deflection channel (d), recorded simultaneously with the cAFM image, show the same wave-like pattern as the scan in (a). A diamond-coated AFM-tip (DEP01) is used. A voltage of 3 V is applied to the back of the sample, while the tip is grounded.

Line patterns, identical to the ones seen in the current channel, are also observed in the height channel of the cAFM scan, displayed in Fig. 1(c). The pattern represents a modulation on top of the height difference that is commonly detected between $+P$ and $-P$ domains on the polar surface of chemo-mechanically polished hexagonal manganite crystals[16–18]. Furthermore, we find that the pattern is readily resolved in the so-called deflection channel (Fig. 1(d)). The deflection signal is recorded simultaneously with the cAFM and height signals while scanning the probe across the electrically biased sample in contact mode; it is thus sensitive to variations in topography and surface potential. Interestingly, in the deflection signal, the wave-like pattern is also observable if no voltage is applied during the scan. However, we cannot fully exclude a potential difference between the tip and the insulating sample surface due to a floating potential, which can give rise to significant electrostatic tip-sample interactions.[19,20] Most importantly for this work, monitoring the pattern in this channel allows to scan the sample multiple times at the same position, offering a less invasive way of imaging compared to scanning with a biased sample and reducing the risk of altering the material's electronic (surface) structure while scanning.[21,22] Therefore, we apply this method for collecting angle-dependent data as displayed in Fig. 2.

### III. ANGLE AND PIXEL-DENSITY DEPENDENCE

To determine whether the patterns observed in Fig. 1 are an intrinsic or extrinsic moiré effect, we perform multiple scans varying the scan angle and density of measurement points. While patterns caused by an intrinsic moiré effect do not depend on these parameters, scanning moiré fringes are co-determined by the scan itself and are thus expected to change shape and periodicity when the scan parameters are varied.

We first perform scan-angle-dependent scans, systematically changing the scan direction with respect to the sample. Fig. 2(a)-(c) shows the deflection signal, recorded with varying scan angles as illustrated in Fig. 2(d)-(f). The data reveals strong variations in the periodic pattern. The change of the fringes manifests itself in a variation of the periodicity of the lines as well as a qualitative change in shape. The latter is strikingly visible in the domain marked with the dotted white oval in Fig. 2(a), where the closed ring structure of the pattern in Fig. 2(b) transforms into a wave-like pattern in Fig. 2(a) and (c).

By analyzing the periodicity of the pattern, we observe two maxima occurring at a scan direction approximately 55° and 145° relative to the cantilever orientation as shown in Fig. 2(g). The periodicity of the wave pattern is measured by counting the maxima in deflection along a line perpendicular to the fringes in the domain marked with a dashed white circle in Fig. 2(a). The data reflects a four-fold symmetry, which is consistent with the symmetry of the scan pattern. The latter can be approximated by a quadratic point pattern as visualized by the red dots in Fig. 2(d)-(f), where a rotation by 90° corresponds to a symmetry operation.

The dependence on the scan angle leads us to the conclusion that the measured pattern is co-determined by the measurement we are performing. The data in Fig. 2 suggests that the pattern results from a convolution of the periodic scan patterns and intrinsic electronic features that affect the conductance of the sample, identifying it as a scanning moiré effect.

To further study the impact of scanning parameters on the observed moiré fringes, we next investigate the influence of the density of measurement points on the pattern formation. Fig. 3 shows cAFM data recorded on another Er(Mn,Ti)O$_3$ specimen cut from the same batch. Here, similar wave-like patterns arise in the scanned area, confirming that the pattern formation is independent of the imaged area or local electronic properties associated with a specific sample.

Fig. 3(a)-(c) shows a cAFM scan of the same area for different pixel densities. By comparing the scans in Fig. 3 (a)-(c), it is clear that the pattern changes periodicity and orientation as the number of pixels is changed in the scan.

Similarly, the density of measurement points can be changed by changing the size of the scan, while keeping the number of measurement points constant. We note that the same effect can also be achieved after scanning by downsampling a high-resolution image and selecting only certain measurement points (see supplementary Fig. S1). Fig. 3 thus shows that the sampling has a substantial impact on the pattern formation, corroborating that it originates from an extrinsic scanning moiré effect.

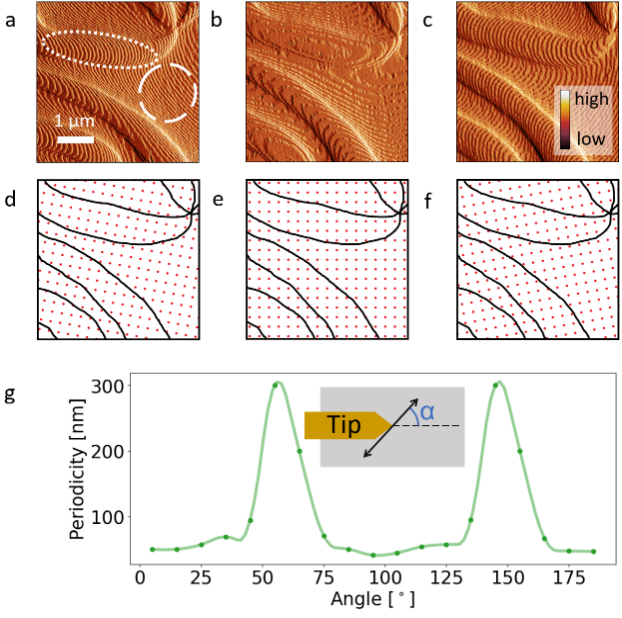

Figure 2. **Angular dependence of wave-like patterns in the deflection signal.** (a)-(c): Deflection signal recorded without bias voltage for different scan angles (135°, 145°, 155° with respect to the fast scan direction). The images are rotated in post-processing to show the same sample area. (d)-(f): Schematic representation of the scan pattern. The red dots represent the measurement points. Note that the red dots do not represent the actual density of scanning points, which was 512 × 512 pixels for the presented data set. (g) Periodicity of the observed pattern as a function of the scan angle with respect to the cantilever orientation (see insert), evaluated in the domain marked with a dashed line in (a). The periodicity was measured by counting the maxima per distance along a line perpendicular to the wave fronts. Due to the change in the shape of the patterns, this line was chosen slightly differently for each angle to ensure that it was perpendicular to the wave patterns. The images are recorded with a diamond-coated DEP01 tip and no bias voltage applied to the sample.

## IV. MICROSCOPIC ORIGIN

After clarifying that the detected moiré fringes are co-determined by the scanning parameters, which identifies the patterns as a scanning moiré effect, we now discuss possible properties of the sample that contribute to this interference phenomenon. When performing cAFM scans with a higher density of measurement points (see Fig. 4(a)), we find an additional periodic line pattern with varying conductance, which has a much smaller periodicity than the emergent moiré fringes discussed so far. To characterize this additional finer pattern, we focus on a smaller area including a ferroelectric domain wall as presented in Fig. 4(b), clearly showing a line pattern of smaller periodicity smoothly continuing over the domain wall with only minor changes in orientation and no visible changes in periodicity. This is qualitatively different from the moiré fringes discussed before, which tend to change their orientation and shape at the domain walls (see, e.g., Fig. 1(a)). The stripe pattern is also visible as steps in the topography when scanning the sample in AC mode (Fig. 4(c)). It has a periodicity of about 25 nm (see Fig. 4(d)), which is comparable to the distance between two scan points in Fig. 1(a) (10 μm × 10 μm, 512 × 512 pixels), which is about 19.5 nm. Importantly, these periodic lines that arise in addition to the moiré fringes are observed within the whole scanned area, and they do not change shape or orientation when changing the scan angle (see supplementary Fig. S2).

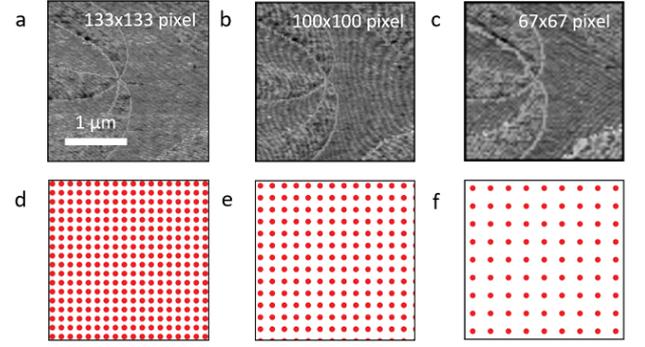

Figure 3. **Influence of the pixel density on the moiré pattern.** (a)-(c) cAFM scans of an Er(Mn,Ti)O$_3$ sample with the polarization perpendicular to the sample surface. The images are recorded with different densities of measurement points as indicated in (d)-(f). In order to make the pattern equally visible in both domains, the images are displayed in grayscale and the current values in both domains are adjusted to the same level in post-processing. In (d)-(f), a dot pattern is presented, representing the scan points in the images (a)-(c). For clarity, not all scan points are displayed, but the ratio between the number of points in the pattern equals the ratio between the number of pixels used to record the images. A diamond-coated tip (DEP01) is used and a bias voltage of 4 V is applied to the back of the sample, while the tip is grounded.

To gain insight into the microscopic origin of the cAFM and topography signals in Fig. 4(b) and (c), we perform high-angle annular dark-field scanning transmission electron microscopy (HAADF-STEM) measurements on a cross-sectional specimen cut from an Er(Mn,Ti)O$_3$ crystal as presented in Fig. 4 (e). The HAADF-STEM data is recorded viewing along the [100] zone axis of Er(Mn,Ti)O$_3$ and reflects the typical layered structure of the material. The Er atoms are visible as the brighter dots in the image, separated by Mn layers, which exhibit a lower contrast. For additional information on the atomic-scale structure, the interested reader is referred to, e.g., ref.[23–25]. The observed surface structure is consistent with a slight misorientation of the crystal surface of about 2.5°. Such a misorientation naturally arises when single crystals are cut and polished into a specific orientation and is difficult to avoid.

Fig. 4(e) reveals that the misorientation leads to a step-like surface structure. Although the cross-sectional data do not allow quantifying the exact step-width and local termination, we estimate an average distance of about 10 nm between the visible Er terminated steps, as indicated by the white arrows. This value is expected to vary for different regions and from sample to sample. The length scale on which it occurs is in the tens of nm regime, similarly to the steps observed in topography (Fig. 4(c)), giving a possible microscopic explanation for the observed surface structure, which may also affect the transport properties. As visible in Fig. 4(d), the periodic variation in the conductance and topography exhibit the same frequency but are out of phase by 180°, which suggests a one-to-one correlation between the measured transport properties and surface structure. Possible sources are variations in surface termination that can arise at misoriented surfaces as schematically shown in Fig. 4(f) or changes in the tip-surface contact area. Most importantly for this work, the results show that nanoscale variations in topography/conductance can lead to pronounced μm-scale moiré fringes in large-scale cAFM scans.

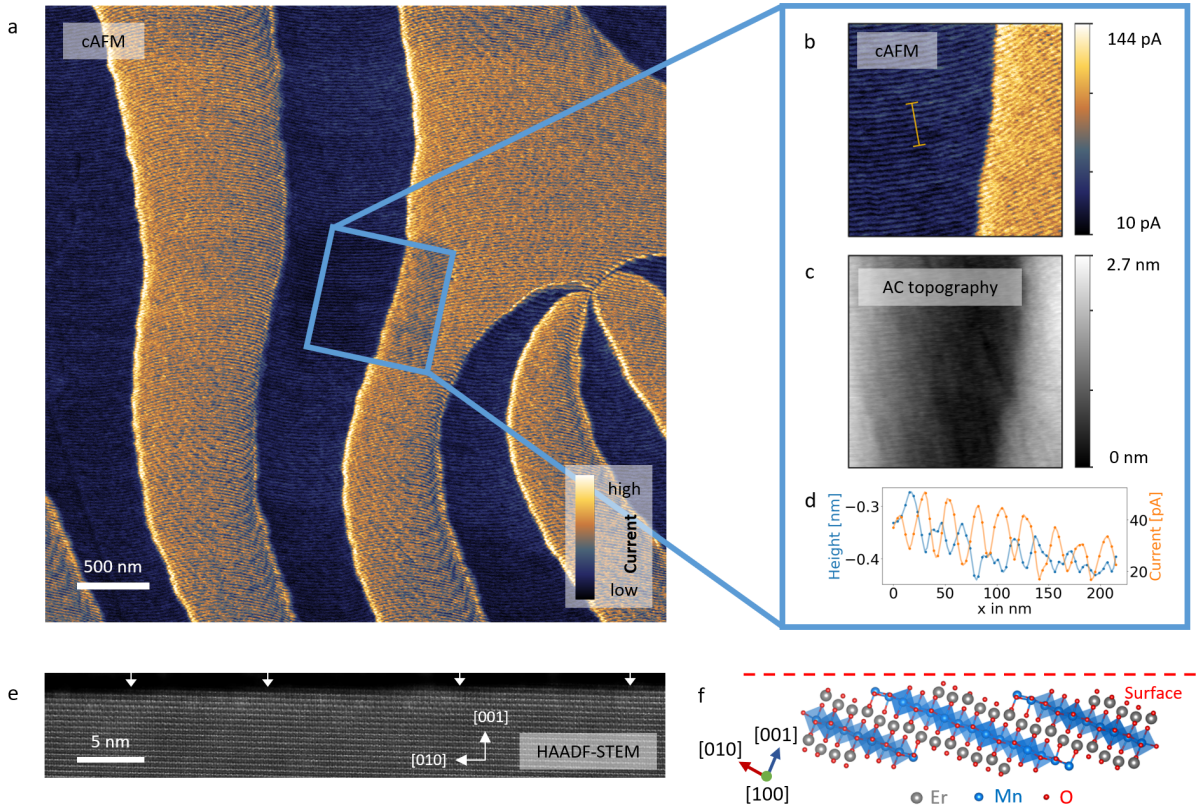

Figure 4. **High-resolution cAFM scans.** (a) High-resolution cAFM scan (1536 × 1536 pixels) of an Er(Mn,Ti)$O_3$ sample with out-of-plane polarization. A periodic stripe-like pattern is visible in addition to the moiré fringes. (b) cAFM scan (256 × 256 pixels) of the region marked in blue in (a). The periodic stripe pattern is clearly visible. The cAFM scans are performed with a diamond-coated tip (DEP01) and 4 V scan voltage. (c) Height profile of an AC topography scan at the same position as (b). The same stripes as in (b) are visible. (d) Current and height profile along the orange line in (b), revealing a periodicity of approximately 25 nm. (e) HAADF-STEM data of a cross-section cut from a different out-of-plane polarized Er(Mn,Ti)$O_3$ sample. (f) Sketch of the crystal structure at a surface with a misorientation of the c-axis of approximately 30° with respect to the surface normal.

## V. SUMMARY

We report the observation of moiré fringes in cAFM conductance maps gained on the ferroelectric semiconductor Er(Mn,Ti)$O_3$. The emerging patterns are observable in the current, deflection, and height channels of the scans and change shape and orientation depending on the scan parameters (scan angle and density of measurement points). Based on these findings, we conclude that the measured moiré patterns are extrinsic in nature, i.e., arise from a superposition of physical properties intrinsic to the sample and the applied raster scanning, classifying them as a scanning moiré effect. A possible candidate for the sample-intrinsic contribution is the step-like surface morphology of the material and its potential impact on the measured conductance.

Our study reveals that subtle variations in the physical properties close to the resolution limit of standard cAFM scans can lead to the formation of moiré patterns in conductance maps with a much larger periodicity. These scanning moiré patterns are extrinsic in nature and a careful experimental characterization is required to distinguish them from intrinsic moiré phys-ics that relate to the material under investigation, independent of the applied probe. Furthermore, analogous to moiré fringe methods in STEM and AFM, the observation of moiré fringes in cAFM opens a possible pathway for improving the sensitivity and range of application for this technique, probing conductance variations at the spatial resolution limit via the analysis of emergent more long-ranged moiré patterns.

## ACKNOWLEDGMENTS

The authors thank J. Masell and K. Shapovalov for fruitful discussions and valuable input. D.M. thanks NTNU for support through the Onsager Fellowship Program and the Outstanding Academic Fellow Program. D.M., L.R., U.L., and J.H. acknowledge funding from the European Research Council (ERC) under the European Union's Horizon 2020 Research and Innovation Program (Grant Agreement No. 863691). The Research Council of Norway is acknowledged for the support to the Norwegian Micro- and Nano-Fabrication Facility, NorFab, project number 295864 and the Norwegian Center for Transmission Electron Microscopy, NORTEM (197405).

## DATA AVAILABILITY STATEMENT

The data that support the findings of this study are available from the corresponding author upon reasonable request.

# Supplementary

## Downsampling of Images

The density of measurement points can be changed in post-processing. This is done by displaying only every *n*-th pixel of the raster scan. Figure S1 displays a comparison between images that were retrieved by downsampling an image with a scan resolution of $500 \times 500$ pixels in Fig. S1(a) to $166 \times 166$ pixels in Fig. S1(b) and to $125 \times 125$ pixels in Fig. S1(c). For comparison, also scans recorded with the respective pixel density are shown (Fig. S1(d) and (e)).

It is visible that the downsampling of the image resolution affects the shape and periodicity of the moiré fringes. Comparing the as-measured images in Fig. S1(d) and (e) and the downsampled images in Fig. S1(b) and (c) shows that the fringes undergo very similar changes in both cases. We can thus conclude that both methods yield the same results, which is expected for a scan sensitive formation mechanism.

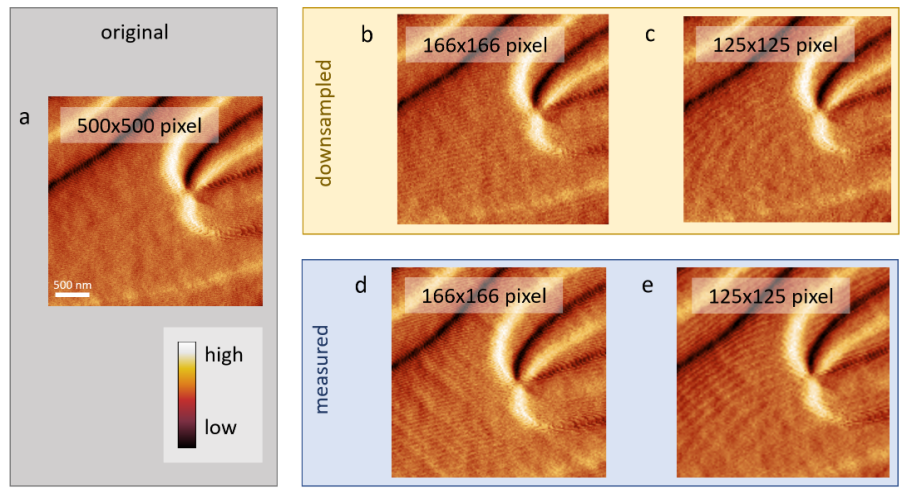

Figure S1: Resulting images from artificially downsampling the deflection channel of a cAFM scan with 0 V bias voltage, performed with a diamond coated DEP01 tip. (a) Original image recorded with $500 \times 500$ pixel. (b) Resulting image when only every $3^{\text{rd}}$ pixel of the scan in (a) is displayed in both directions, corresponding to a resolution of $166 \times 166$ pixels. (c) Resulting image when only every $4^{\text{th}}$ pixel of the scan in (a) is displayed in both directions, corresponding to a resolution of $125 \times 125$ pixels. (d) and (e) scans with lower resolution (position slightly shifted with respect to (b) and (c)).

## High-periodicity stripe patterns

To further investigate the origin of the fine stripe pattern observed in Fig. 4, we perform additional measurements. Fig. S2 shows two scans of the stripe pattern, with a difference in scan direction of 10° between them. No difference in the stripe pattern is visible, excluding a scanning moiré effect as its cause.

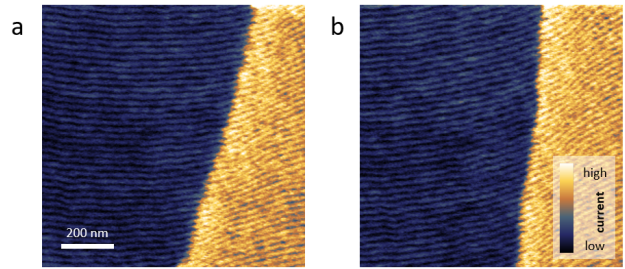

Figure S2: cAFM scans of the high-periodicity stripe pattern discussed in Fig. 4 in the main text. (a) is recorded with a scan direction of 20° and (b) with a scan direction of 30°. The scan resolution is 256 × 256 pixels and the cAFM scan is done with a diamond-coated DEP01 tip and a bias voltage of 4 V.

To exclude periodic noise as the source of the observed periodic stripe patterns, we perform scans at different scan speeds. If the pattern was caused by periodic noise, this would result in a change in the pattern's periodicity and orientation. As visible in Fig. S3, this is not the case.

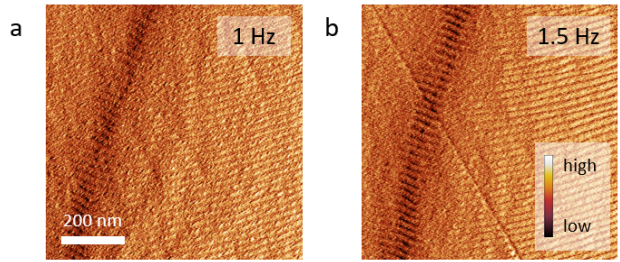

Figure S3: Deflection signal of the cAFM scans of the high-periodicity stripe pattern discussed in Fig. 4 in the main text. (a) is recorded with a scan speed of 1 Hz and (b) with a scan speed of 1.5 Hz. The scan resolution is 424 × 424 pixels and the cAFM scan is done with a diamond-coated DEP01 tip and a bias voltage of 3 V. The scan positions in (a) and (b) are slightly shifted. The diagonal line in (b) is caused by the tip movement during the AFM scan in (a).